\documentclass[12pt]{iopart}
\usepackage{iopams}  
\usepackage{graphicx}
\usepackage[pdfborder={0 0 0}
						,bookmarks=false				
            ]{hyperref}
\usepackage[numbers]{natbib}  

\newcommand{\OOp}{O$_2^+$} 
\newcommand{\tfrac}[2]{\textstyle\frac{#1}{#2}} 
\newcommand{\newblock}{\ }

\begin{document}
\title[Optical clocks based on molecular vibrations as probes of $\mu$-variation]{Optical clocks based on molecular vibrations as probes of variation of the proton-to-electron mass ratio}

\author{David Hanneke, Boran Kuzhan and Annika Lunstad}

\address{Physics \& Astronomy Department, Amherst College, Amherst, Massachusetts 01002, USA}
\ead{dhanneke@amherst.edu}

\begin{abstract}
Some new physics models of quantum gravity or dark matter predict drifts or oscillations of the fundamental constants.
 A relatively simple model relates molecular vibrations to the proton-to-electron mass ratio $\mu$. Many vibrational transitions are at optical frequencies with prospects for use as highly accurate optical clocks.
 We give a brief summary of new physics models that lead to changes in $\mu$ and the current limits on drifts and oscillation amplitudes.
 After an overview of laboratory procedures, we give examples of molecules with experiments currently in development or underway. 
 These experiments' projected systematic and statistical uncertainties make them leading candidates in next-generation searches for time-variation of $\mu$.
\end{abstract}

\vspace{2pc}
\noindent{\it Keywords}: optical clocks, molecular clocks, fundamental constants, proton-to-electron mass ratio, variation of constants, molecular vibration

\section{Introduction}

Atoms and molecules are compact, naturally occurring systems that can be isolated from their environment in relatively small numbers. Many pairs of quantum states within an atom or a molecule have transition energies corresponding to light in radio through optical frequencies that are conveniently produced in a laboratory. Atomic and molecular transitions are the most accurate frequency references in existence~\cite{riehleMetrologia2018}. They define the SI second and will serve as next-generation clocks based on optical transitions~\cite{ludlowRMP2015}. Because of their high accuracy, atomic and molecular transitions are sensitive probes of physics typically associated with energy scales much higher than the transition energy itself. They serve to demonstrate and test known physics, such as parity violation and quantum electrodynamics, as well as to search for new physics, such as time-reversal violation, quantum gravity, and dark matter~\cite{safronovaRMP2018}.

Some new physics models predict changes to the values of fundamental constants, which are typically referred to in terms of dimensionless values such as the fine structure constant $\alpha=e^2/(4\pi\epsilon_0\hbar c)$ or the proton-to-electron mass ratio $\mu=m_p/m_e$. These changes could take the form of slow drifts or steady oscillations. In this manuscript, we focus on the use of molecules to search for variation in the proton-to-electron mass ratio. Unlike atoms, molecules have direct sensitivity to changes in $\mu$. Their vibrational and rotational degrees of freedom involve the motion of the inertial masses of the nuclei, and the strength and length of their chemical bonds involve the electron mass.

The mass ratio $\mu$ is sensitive to a variety of new effects because the proton and electron acquire their masses by different mechanisms. Protons, neutrons, and their combinations in atomic nuclei are bound states of quarks and gluons. Around 99\,\% of their mass arises from the strong interaction. Each mass is proportional to the quantum chromodynamics energy scale $\Lambda_{\rm QCD}$, ignoring the small contribution from the quark masses~\cite{uzanLRR2011}. This common proportionality is why we talk about molecular vibration and rotation as being sensitive to the proton mass when the literal sensitivity is to the reduced mass of the nuclei. By contrast, the electron is a fundamental particle. Its mass arises from its interaction with the Higgs field. Thus, $\mu$ is sensitive to new physics that couples to gluons, quarks, electrons, or the Higgs.

\section{Changing constants}

\subsection{Models}
Many theories of new physics beyond the Standard Model give rise to dynamical ``constants''. These theories often postulate new scalar fields or extra spatial dimensions. Couplings to standard particles will cause our (3+1)-dimensional constants to change with the dynamics of the new fields or the expanding or contracting length scales of the new dimensions. Several models are reviewed in~\cite{uzanLRR2011}. In many of these models, a common underlying mechanism causes correlations in the values of the fundamental constants as they change. For example, some models predict fractional changes in $\Lambda_{\rm QCD}$ (and thus $m_p$ and $\mu$) to be around 40 times larger than changes in $\alpha$. 
Many models link the constants to cosmological evolution and thus predict drifts over long timescales. They typically involve new massless fields.

By contrast, new massive fields can induce oscillations in the fundamental constants~\cite{arvanitakiPRD2015,stadnikPRL2015,heesPRD2018}. Bosonic matter that is lightweight enough that the particle wavepackets overlap forms a coherent classical wave. The field $\phi$ oscillates at the Compton frequency $f_\phi = m_\phi c^2/h$, where $m_\phi$ is the mass of the new particle. The oscillation amplitude $\phi_0$ is related to the field's energy density $\rho_\phi$ through 
\begin{equation}
\phi_0=\sqrt{\frac{\hbar^{3}}{c^5}}\frac{\sqrt{2\rho_\phi}}{m_\phi}.
\end{equation}
(Here, we write the field amplitude with units of mass, as in~\cite{arvanitakiPRD2015}; see~\cite[App.\,A]{heesPRD2018} for different conventions.) Since it is massive, this field could potentially make up a significant fraction of the galaxy's dark matter; if it does, then this energy density is $\rho_{\rm DM}\approx 0.4~{\rm GeV}/{\rm cm}^3$~\cite{catenaJCAP2010}. New couplings to Standard-Model particles would transfer oscillations in the dark matter field amplitude to effective oscillations in the fundamental constants. In a generic model, the new couplings to each Standard-Model particle are independent, but the fundamental constants would oscillate with a common frequency and phase. In some specific models, such as the relaxion model~\cite{grahamPRL2015,banerjeePRD2019}, the couplings to Standard-Model particles are related to each other as well. In these cases, there is a model-specific relation between the oscillation amplitudes of the various fundamental constants. The allowed dark matter particle masses span many orders of magnitude, from $m_\phi\sim10^{-21}~{\rm eV}/c^2$ (the de~Broglie wavelength should fit in the halo of the smallest dwarf galaxies) to $10^{-1}~{\rm eV}/c^2$ (the density of particles in the local dark matter should be sufficient to form a classical field)~\cite{stadnikPRL2015,stadnikPRA2016}. This mass range corresponds to oscillation frequencies in the range $f_\phi\sim10^{-7}$--$10^{13}$~Hz.

\subsection{Current Limits}

The tightest constraints on these models come from direct searches for drifts of atomic and molecular spectral lines as well as indirect limits on oscillations from equivalence-principle tests. Observations of molecular spectra in high-redshift astrophysical scenarios constrain 
\begin{equation}
	{|\Delta\mu|}/{\mu} < 10^{-6}-10^{-7} \quad {\rm over}\quad \sim 10^{10}~{\rm yr}.
\end{equation}
Comparisons with present-day spectra have been made for a variety of species, including NH$_3$, CS, and H$_2$CO~\cite{kanekarApJL2011}, CH$_3$OH~\cite{bagdonaitePRL2013}, H$_2$~\cite{ubachsRMP2016}, and CO~\cite{dapraMNRAS2017}.

The tightest constraint on present-day drift comes from comparison of an Yb optical clock with a Cs microwave clock~\cite{mcgrewOptica2019}, which set the limit
\begin{equation}
	{\dot{\mu}}/{\mu} = \left(-5.3\pm 6.5\right)\times10^{-17}~{\rm yr}^{-1}.  
\end{equation}
This limit is an improvement by a factor of approximately two over results from several years earlier that compared an optical clock based on the Yb$^+$ ion with Cs~\cite{godunPRL2014,huntemannPRL2014}. In each of these cases, the optical transition is electronic in nature and has almost no sensitivity to changes in $\mu$. The hyperfine microwave transition's $\mu$-dependence is in the cesium nuclear magnetic dipole moment. The relationship of this moment to $\mu$ is model-dependent~\cite{flambaumPRC2006}.

Optical clocks have surpassed microwave clocks in accuracy~\cite{ludlowRMP2015}. For this reason, next-generation experiments are focusing on comparing two optical transitions. Molecules possess optical transitions with sensitivity to $\mu$ because their vibrational and rotational degrees of freedom involve the motion of the nuclei themselves. Molecular clocks are not yet at the accuracy of atomic clocks, though many molecular species have the potential to rival atomic clocks in terms of their ultimate systematic effects. The current best limit on $\mu$-drift that uses a molecule is based on two vibrational states in different electronic potentials in KRb. The limit set is~\cite{kobayashiNatureComm2019}
\begin{equation}
 \dot{\mu}/\mu = (-0.30\pm1.0)\times10^{-14}~\textrm{yr}^{-1}. 
\end{equation}
Earlier experiments used an optical transition in a beam of SF$_6$ to set a six-times less stringent limit~\cite{shelkovnikovPRL2008}.

We are unaware of any experimental results that directly probe for oscillations in $\mu$. We can estimate limits on $\mu$ oscillation amplitudes from indirect probes by assuming a phenomenological model of the dark-matter coupling. Fundamental constants that change in time would appear to violate the equivalence principle; prior experiments searching for equivalence-principle violations provide constraints on the dark matter coupling coefficients~\cite{arvanitakiPRD2015,heesPRD2018}. These constraints tend to be $m_\phi$-independent up to some threshold mass. For example, torsion balance experiments~\cite{schlammingerPRL2008} restrict the coupling parameter $|d_g|<7.2\times 10^{-6}$ for masses whose Compton wavelength is greater than the Earth's radius ($m_\phi<3\times10^{-14}~{\rm eV}/c^2$). Here, $d_g$ is a dimensionless parameter characterizing the strength of a linear coupling of $\phi$ to the gluon field; a value of one would represent a force of the same strength as gravity. 
 At higher masses, the restriction relaxes to approximately $|d_g|<3\times10^{-2}$. See~\cite{arvanitakiPRD2015} for a discussion and more detailed exclusion plots; see~\cite{heesPRD2018} for a discussion of a phenomenological model with quadratic coupling $\phi^2$. 
If we assume the only coupling is to the gluon field, the oscillation amplitude of the fractional change in $\mu$ is given by~\cite{arvanitakiPRD2015,heesPRD2018}
\begin{equation}
	\frac{\Delta\mu}{\mu} = d_g\phi_0 {\sqrt{\frac{4\pi G}{\hbar c}}} = \frac{d_g\sqrt{\rho_\phi}}{m_\phi}\frac{\hbar\sqrt{8\pi G}}{c^3}.
\end{equation}
Here, $G$ is Newton's gravitational constant. Assuming further that the new field's energy density is equal to the local dark matter density, the equivalence principle limits correspond to an oscillation amplitude of approximately
\begin{equation}
	\frac{|\Delta\mu|}{\mu} < \left\{\begin{array}{l} 5\times10^{-36}~{\rm eV}/(m_\phi c^2) , \qquad m_\phi < 3\times10^{-14}~{\rm eV}/c^2 \\
								 2\times10^{-32}~{\rm eV}/(m_\phi c^2) , \qquad m_\phi > 3\times10^{-14}~{\rm eV}/c^2\end{array}\right. .
\end{equation}
In terms of the dark matter's Compton frequency, which shows up as the experimental modulation frequency, these limits correspond to
\begin{equation}
	\frac{|\Delta\mu|}{\mu} < \left\{\begin{array}{l} 1\times10^{-21}~{\rm Hz}/f_\phi , \qquad f_\phi < 7~{\rm Hz} \\
								 5\times10^{-18}~{\rm Hz}/f_\phi , \qquad f_\phi > 7~{\rm Hz}
								\end{array}\right. .
\end{equation}
For example, at an oscillation frequency of 1~mHz ($m_\phi=4\times10^{-18}~{\rm eV}/c^2$), $\mu$ is already constrained to oscillate with an amplitude below $\sim10^{-18}$. At 1~kHz ($m_\phi=4\times10^{-12}~{\rm eV}/c^2$), the constraint is $\sim5\times10^{-21}$. Since the oscillation amplitude scales as $m_\phi^{-1}$, the low-mass/low-frequency end of the range is the place where molecular clocks have the greatest chance to improve over equivalence-principle limits.

There exist models where the dark matter can become bound to the Earth or Sun such that it would have substantially larger densities than the galactic average~\cite{banerjeeCommunPhys2019}. The ensuing larger oscillation amplitudes would be most pronounced if $m_\phi$ is within a few orders of magnitude of $10^{-10}~{\rm eV}/c^2$ ($f_\phi=10^4$~Hz) for dark matter bound to the Earth and $10^{-15}~{\rm eV}/c^2$ ($10^{-1}$~Hz) if bound to the Sun. In these scenarios, molecular experiments have better prospects for reaching the limits set by equivalence-principle tests.

\section{Gaining sensitivity to constants}\label{sec:sensitivity}

\subsection{Sensitivity in general}
A search for time-variation in $\mu$ begins by choosing two quantum states that depend differently on the mass ratio. For a single state, we can write an absolute sensitivity factor $q_\mu$ that quantifies the absolute energy shift $\Delta E$ for a fractional change in $\mu$:
\begin{equation}
	\Delta E = q_\mu \frac{\Delta\mu}{\mu}.  
\end{equation}
An experiment requires two states with different sensitivity factors. One then monitors the transition frequency
\begin{equation}
 hf = E^\prime-E^{\prime\prime}
\end{equation}
looking for changes
\begin{equation}
	h\,\Delta f = \Delta E^\prime - \Delta E^{\prime\prime} = (q^\prime_\mu-q^{\prime\prime}_\mu)\frac{\Delta \mu}{\mu}.
\end{equation}
For present-day searches, we often want as large a frequency shift as possible. To compare the absolute shift of various transitions, we define the absolute sensitivity of a transition:
\begin{equation}
	f_\mu \equiv \mu\frac{\partial f}{\partial\mu} = \frac{q^\prime_\mu-q^{\prime\prime}_\mu}{h},
\end{equation} 
such that
\begin{equation}
	\frac{\Delta \mu}{\mu} = \frac{\Delta f}{f_\mu}. \label{eq:fmu}
\end{equation}
This quantity gives the absolute frequency shift of the transition for a fractional shift in $\mu$. For example, if $f_\mu = 100~{\rm THz}$, then a shift in $\mu$ at $10^{-16}$ would give a 10~mHz shift in the transition frequency.

In the literature, one often encounters the relative sensitivity factor $K_\mu$
\begin{equation}
	\frac{\Delta\mu}{\mu} = \frac{1}{K_\mu} \frac{\Delta f}{f},\label{eq:Kmu}
\end{equation}
which is sometimes called the relative enhancement factor. It is related to the absolute sensitivity by $K_\mu = f_\mu/f$. The relative sensitivity is useful in contexts where the precision with which one can measure a frequency shift scales with the frequency. An example is the Doppler-broadened lines common in astrophysical data. In a laboratory context, relative sensitivity can be useful in cases of accidental degeneracies of levels with different absolute sensitivities. In these cases, it might be possible to achieve optical sensitivities in $f_\mu$ with measurements done with $f$ in the microwave. The experimental enhancement comes only if one can directly measure the frequency difference between the two states. Appropriate near-degeneracies have been identified in Cs$_2$~\cite{demillePRL2008} and \OOp~\cite{hannekePRA2016}, and used for a $\dot{\mu}$ measurement in KRb~\cite{kobayashiNatureComm2019}.

In applying these sensitivity factors to oscillating constants, the molecular and dark-matter coherence times need to be considered. The dark matter is expected to be moving at the galactic virial speed $v\sim10^{-3}c$~\cite{kraussPRL1985}, which would lead to oscillation quality factors of order $Q\sim10^6$ and coherence times $\tau_{\rm coh}\sim 10^6/f_\phi$~\cite{stadnikPRL2015a,arvanitakiPRD2015,dereviankoPRA2018}. The duration of a single measurement $T_m$ should be shorter than $\tau_{\rm coh}$ so that the signal does not average to zero. If the total duration $\tau$ of multiple measurements is also less than $\tau_{\rm coh}$, then the dark matter's phase coherence can be used to maintain sensitivity to $\mu$ variation~\cite{arvanitakiPRD2015,dereviankoPRA2018,degenRMP2017}. 
  The sensitivity is suppressed for measurement durations longer than $\tau_{\rm coh}$ or experiments where the dark matter phase is allowed to vary randomly for each individual measurement~\cite{dereviankoPRA2018,kotlerPRL2013,degenRMP2017}.

\subsection{Sensitivity in molecules}
The scaling of molecular energies with $\mu$ is well-known and has long been used to explain the spectra of isotopologues~\cite[III.2.g]{herzbergVolI1950}. A low-order expansion of molecular energy with vibrational quantum number $v$ and total angular momentum $J$ is
\begin{equation}
E/(hc) = T_e + \omega_e(v+\tfrac{1}{2})-\omega_ex_e(v+\tfrac{1}{2})^2+B_eJ(J+1),
\end{equation}
where the expansion coefficients are given in the traditional units of wavenumbers. The electronic energy $T_e$ is independent of $\mu$. The vibrational coefficient $\omega_e$ scales as $1/\sqrt{\mu}$ (that is, $d\omega_e/\omega_e = -\tfrac{1}{2}\, d\mu/\mu$). The lowest anharmonicity coefficient $\omega_ex_e$ and the rotational constant $B_e$ each scale as $1/\mu$. The scaling of coefficients in higher-order expansions may be found in~\cite{herzbergVolI1950,watsonJMS1980,beloyPRA2011,pastekaPRA2015}. Given these scalings, the absolute sensitivity of this molecular state is
\begin{equation}
	q_\mu = hc\left[-\tfrac{1}{2}\omega_e(v+\tfrac{1}{2})+\omega_ex_e(v+\tfrac{1}{2})^2-B_eJ(J+1)\right].
\end{equation}

\begin{figure}
\begin{center}
\includegraphics[width=0.7\textwidth]{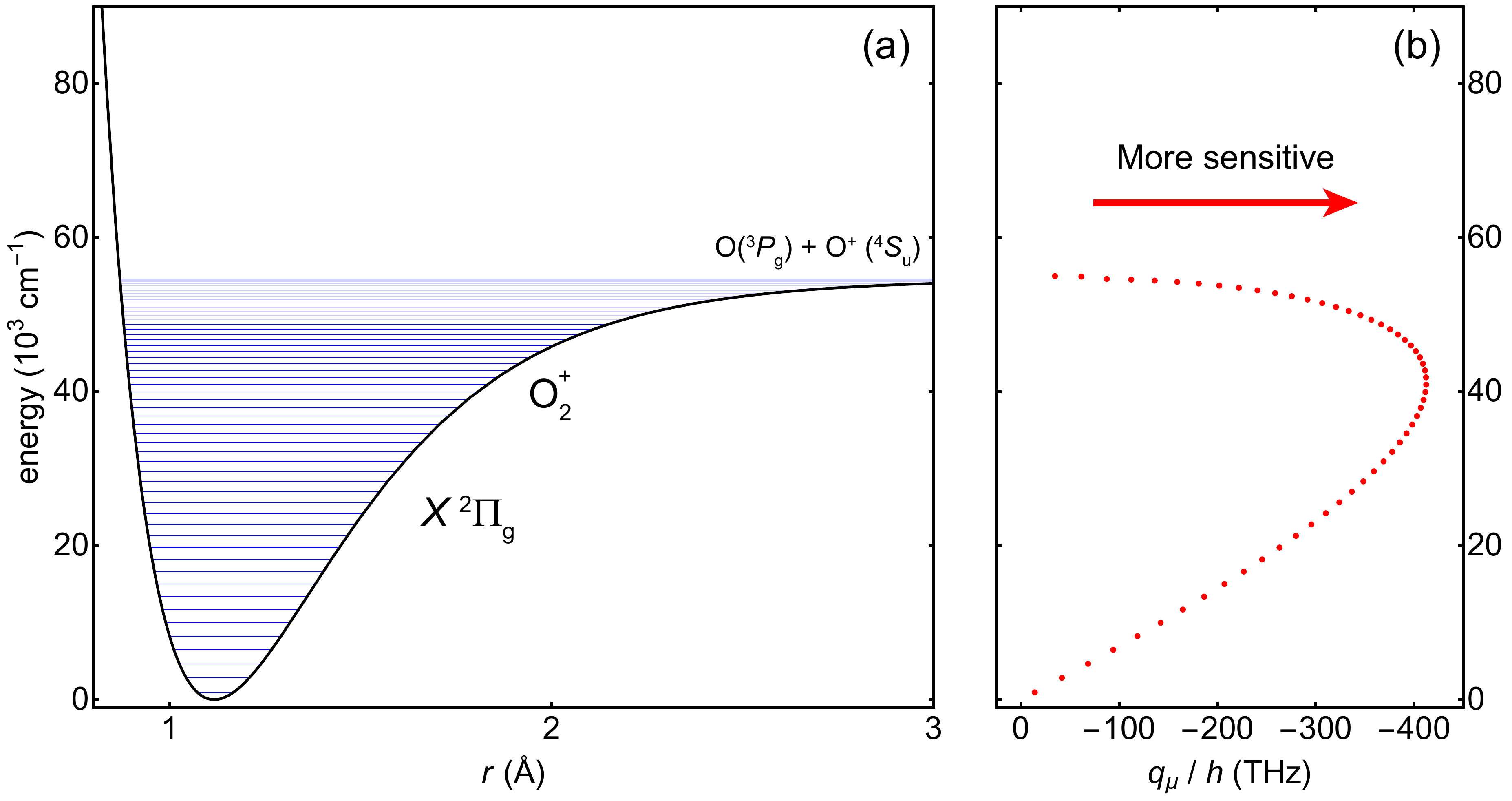}
\caption{Vibrational levels and absolute sensitivity for the \OOp~ground $^2\Pi_g$ state. (a) Potential curve. Horizontal lines indicate the energies of vibrational states. Dark lines indicate levels that have been observed experimentally~\cite{songJCP1999}. Light lines fill in the calculated energies of the remaining levels~\cite{carolloAtoms2018}. (b) Absolute sensitivity of each vibrational state. The values are negative because larger $\mu$ would reduce the vibrational energies.}\label{fig:sensitivity}
\end{center}
\end{figure}

Vibrational transitions hold the potential for the highest sensitivity. The vibrational coefficient is typically much larger than the rotational coefficient, and angular momentum conservation prevents transitions with large $\Delta J$. For small $v$, $q_\mu$ increases linearly with $v$; vibrational transitions within the harmonic part of the potential have relative sensitivity $K_\mu = -\tfrac{1}{2}$. Higher in the potential, the anharmonic term reduces the absolute sensitivity, causing it to head back toward zero near dissociation. This reduction makes sense, as near dissociation the system has properties closer to two atoms than a molecule. The maximum sensitivity occurs for vibrational states of energy approximately $3/4$ of the dissociation energy~\cite{zelevinskyPRL2008,demillePRL2008}. \Fref{fig:sensitivity} shows the sensitivity of vibrational states in the \OOp~molecule, and clearly shows the initial linear growth with $v$ and maximum at $v=28$.

Although this manuscript focuses on variation in $\mu$, some transitions may also have sensitivity to variation in the fine structure constant $\alpha$. This is especially true of any transition involving states with different spin--orbit couplings or different relativistic corrections to their electronic potentials~\cite{flambaumPRL2007,beloyPRA2011a}.

\section{Experiments with molecules}\label{sec:experiments}

\subsection{Probing the transition, state preparation, and measurement}

Most experimental proposals focus on narrow-line vibrational transitions with high absolute sensitivity $f_\mu$. Narrow lines afford longer coherent probe times, and thus faster averaging of statistical uncertainties (\sref{sec:stats}). In many -- but not all -- cases, a nonpolar molecule is chosen so the vibrational transitions within an electronic state are electric-dipole ($E1$) forbidden. They can still be driven as electric quadrupole ($E2$) or two-photon transitions. In these precision measurements, a thorough analysis of systematic effects (\sref{sec:systematics}) also plays an important role. Long probe times are facilitated by trapping the molecules. Ion traps and optical lattices can provide tight confinement that eliminates first-order Doppler shifts. Experiments searching for a drifting $\mu$ probe the relevant transition frequencies multiple times over the course of a year or more and fit a line with slope $\dot{f}$.

To get a sense for how an experiment searching for oscillations in $\mu$ might proceed, we can look to recent work using atoms to search for oscillations in $\alpha$. For oscillations with a period longer than the experiment cycle-time, one can record the result of each experiment and Fourier transform the data~\cite{arvanitakiPRD2015,dereviankoPRA2018}. The idea is similar to the sinusoidal fits already done with atomic clock data when searching for annual modulation of the clock frequency as the Earth moves in the Sun's gravitational well~\cite{blattPRL2008,safronovaRMP2018,mcgrewOptica2019}. For periods shorter than a single experiment, a dynamical decoupling pulse sequence can enhance the sensitivity to AC modulation of the molecular frequency~\cite{kotlerNature2011,kotlerPRL2013,aharonyArxiv2019}. The particular pulse sequence chosen shapes a frequency filter function~\cite{degenRMP2017} and would have to be modified during a search through the wide frequency space. If care is taken to synchronize subsequent dynamical decoupling sequences to the oscillations, the signal can maintain its sensitivity to $\mu$ up to $\tau\sim\tau_{\rm coh}$. If the $\mu$-oscillation phase is allowed to vary for each measurement, the signal has only quadratic sensitivity to the oscillation amplitude~\cite{kotlerPRL2013,degenRMP2017}.

Alternatively, experiments using traditional spectroscopy techniques~\cite{demtroderBook} such as polarization spectroscopy will show direct AC modulation of their signal~\cite{antypasPRL2019,antypasAnnPhys2020}. For example, the signal might be a photodiode's current that corresponds to the transmission though a molecular vapor cell of a certain frequency and polarization of light. Such techniques often rely on steady-state behavior that is only present on timescales much longer than the molecular state spontaneous emission time. For times much shorter than $1/\gamma$ ($\gamma$ being the natural linewidth in non-angular frequency units), non-steady-state behavior impedes the signal from tracking the molecular frequency. Thus, high-frequency modulations of $\mu$ are suppressed in the measured frequency shifts. This measured shift needs to be corrected as~\cite{antypasPRL2019}
\begin{equation}
	\frac{\Delta\mu}{\mu} = \frac{1}{K_\mu} \left(\frac{\Delta f}{f}\right)_{\rm actual}=\frac{1}{K_\mu} \left(\frac{\Delta f}{f}\right)_{\rm measured}\sqrt{1+\left(\frac{f_\phi}{\gamma}\right)^2}. \label{eq:linewidthSensitivity}
\end{equation}

A variety of choices exist for state preparation and measurement. Some state-preparation procedures create the molecule from other resources. Bi-alkali molecules may be prepared by photoassociating them from ultracold atoms~\cite{zelevinskyPRL2008,kobayashiNatureComm2019}. Molecular ions can be photoionized from the corresponding neutral molecule~\cite{tongPRL2010}. Other procedures prepare the state through dissipation. Molecules with near-closed transitions are amenable to optical pumping~\cite{stollenwerkAtoms2018}. Molecular ions co-trapped with atomic ions can use a quantum projection protocol~\cite{chouNature2017}.

State-detection procedures can be classified by whether they destroy the molecule or not. State-selective photodissociation can be followed by photofragment imaging~\cite{kondovNatPhys2019}, by mass analysis in an ion trap or time-of-flight mass spectrometer, or simply by observing molecule loss~\cite{bresselPRL2012}. Another destructive technique is state-selective chemistry, such as laser-induced charge transfer with a background gas~\cite{tongPRL2010}. Optical cycling with fluorescence detection~\cite{stollenwerkAtoms2018} provides non-destructive detection for molecules with diagonal Franck-Condon factors. Quantum-logic spectroscopy~\cite{schmidtScience2005,chouNature2017} is a general-purpose technique for detecting the state of a molecular ion by transferring it to a co-trapped atomic ion that is amenable to fluorescence detection.

A single molecular transition is not sufficient to detect a change in $\mu$. It must be compared to another reference that has different absolute sensitivity and comparable measurement accuracy. For searches for $\mu$ drifts, optical atomic clocks~\cite{ludlowRMP2015} provide a suitable reference. They are usually based on electronic transitions that have little sensitivity to $\mu$ variation. An optical frequency comb can readily compare light at different wavelengths. Alternatively, the same molecule often has many transitions with different $f_\mu$, and two or more transitions could be monitored for differential drifts. Such self-reference could lead to common systematic effects that may improve the ultimate accuracy achieved.

When searching for oscillations in $\mu$, the laser itself may be able to serve as the reference. In optical clock experiments~\cite{ludlowRMP2015}, the short-term stability of the laser comes from referencing it to an optical resonance of a two-mirror Fabry-P\'{e}rot interferometer. The optical resonance frequency is set by the distance between the two mirrors, which is maintained by a rigid spacer such as ultra-low-expansion glass or crystalline silicon. The spacer's length has very little sensitivity to $\mu$. For frequencies below the lowest mechanical resonance (typically of order $10~{\rm kHz}$), the length is sensitive to changes in the Bohr radius and thus in $\alpha$ and the electron mass~\cite{geraciPRL2019,antypasPRL2019,aharonyArxiv2019}. (While the electron mass has dimensions, the sensitivity is to mass changes relative to the average mass; this ratio is dimensionless~\cite{antypasAnnPhys2020}.) For even lower frequencies, the length is subject to thermal drifts and other perturbations.

\subsection{Example molecules}

Most experiments to date have focused on searches for long-term drifts and oscillation periods longer than a single experiment cycle. To give a sense of the variety of approaches underway, we list a sample of molecules with active experimental work.

\textbf{KRb}: The current best drift limit using molecules comes from photoassociated KRb~\cite{kobayashiNatureComm2019}. The experiment exploits an accidental near-degeneracy between the ground $X\,^1\Sigma^+$ state and the overlapping, shallow $a\,^3\Sigma^+$ potential to achieve absolute sensitivities in the terahertz with a microwave experiment ($f_\mu = 9.45(4)~{\rm THz}$, $K_\mu = 14\,890(60)$). This experiment yielded the statistics-limited result: $\dot{\mu}/\mu = (-0.30\pm 1.00_{\rm stat}\pm0.16_{\rm syst})\times 10^{-14}~{\rm yr}^{-1}$. As a heteronuclear molecule, its primary systematic effect is blackbody radiation, which will require additional calculations and calibrations to reach higher accuracy. 

$\textbf{Sr}\mathbf{_2}$: Two-photon Raman transitions in lattice-confined photoassociated Sr$_2$ can have sensitivities of up to $f_\mu \approx -7.5~{\rm THz}$~\cite{zelevinskyPRL2008}. Because of the photoassociation process, the natural choice for such a transition is between a vibrational state near the dissociation limit and one deeper in the potential. These molecules have a $X\,0^+_g$ ground electronic state and no hyperfine structure, so experiments using the $J=0$ rotational state should be largely insensitive to magnetic-field systematic effects. Raman clock transitions have been driven coherently in molecules confined to a 1D optical lattice~\cite{kondovNatPhys2019}. Importantly, the lattice and two-photon lasers were all chosen in a manner to suppress any differential light/AC-Stark shifts of the clock states. Residual light shifts from imperfect control of the Raman beam intensities remain the leading systematic effect. 

$\textbf{N}\mathbf{_2^+}$: This ion has been produced in its rovibrational ground state by resonance-enhanced multi-photon ionization (REMPI) and co-trapped with Ca$^+$ atomic ions~\cite{tongPRL2010}. The lowest vibrational transition ($v=0\rightarrow 1$, $f_\mu=-32~{\rm THz}$) in its $X\,^2\Sigma_g^+$ electronic ground state has been driven as an electric-quadrupole transition~\cite{germannNaturePhys2014}. The molecule's lowest excited electronic state ($A\,^2\Pi_u$) provides a convenient electric-dipole transition. For example, the excited state's $v^\prime=2$ level can be accessed with a laser of wavelength $\sim787$~nm. It has been used to apply a state-dependent optical dipole force and detect the population in the $X$ state's ground rovibrational state by use of a quantum-nondemolition technique~\cite{sinhalScience2020}.

$\textbf{O}\mathbf{_2^+}$: Proposals to use this ion include driving two-photon transitions from the ground vibrational state to states with sensitivities as high as $f_\mu = -398~{\rm THz}$ ($v^\prime = 28$ or $29$)~\cite{carolloAtoms2018} or driving electric-quadrupole transitions as high as $v^\prime = 6$ ($f_\mu = -151~{\rm THz}$)~\cite{wolfArxiv2020} or even higher~\cite{kajitaPRA2017}. The most common isotope of oxygen ($^{16}{\rm O}$, $99.8\,\%$ abundance) has no nuclear spin, and thus the molecule has no hyperfine structure. In our own lab, we have photoionized a supersonically expanded molecular beam of O$_2$ to produce \OOp~in its ground vibrational state.

$\textbf{H}\mathbf{_2^+}$, $\textbf{HD}\mathbf{^+}$: Several experiments exist with high-precision spectroscopy of these simplest of molecules~\cite{bresselPRL2012,biesheuvelNComms2016}. Since it is possible to calculate the theory with high accuracy, a main goal of experiments with molecular hydrogen ions is to determine values of fundamental constants, including $\mu$ itself~\cite{alighanbariNature2020}. They are also viable candidates for measuring $\mu$-variation~\cite{schillerPRA2005,schillerPRL2014}.

$\textbf{TeH}\mathbf{^+}$: This polar molecule's relatively deep potential allows for high-sensitivity dipole-allowed vibrational transitions. For example, $v=0\leftrightarrow v^\prime=8$ has $f_\mu= -170~{\rm THz}$~\cite{kokishPRA2018}. 
It has near perfect overlap of vibrational wavefunctions (diagonal Franck-Condon factors) in its ground $X_1\,0^+$ and excited $X_2\,1$, $a_2$, and $b\,0^+$ states. This feature should allow optical pumping for state preparation and quasi-cycling for state detection, with potential reduction in experiment dead time and corresponding improvement in statistical uncertainty~\cite{stollenwerkAtoms2018}.

\section{Long-term prospects}

There are slightly more than two orders of magnitude between the best limit on $\mu$ drifts set with molecules and those set with atoms. Many molecules show prospects for closing this gap and going beyond. Several of the above molecules are estimated to be able to achieve systematic and statistical uncertainties comparable to or better than current optical atomic clocks. They should be able to reach fractional frequency instabilities at $<10^{-18}$. Fractional limits at this level should be comparable to or better than equivalence-principle limits for $\mu$ modulation by the lightest dark matter particles. Limits at higher modulation frequencies from heavier particles are much more stringent and will be more challenging to match.

\subsection{Systematic uncertainties} \label{sec:systematics}
Molecular clocks are subject to many of the same systematic effects as atomic clocks. These effects are reviewed in~\cite{ludlowRMP2015}. There are several important differences to consider with the extra vibrational and rotational structure of molecules. For example, many of the proposed experiments use two vibrational levels in the same electronic state, such that some systematic effects may be common. The energy difference of adjacent vibrational and rotational states are relatively smaller than electronic states, which can alter the scale of some effects.

Electric-field (Stark) shifts have the largest potential to show this changed magnitude. Eigenstates of an unperturbed molecule do not have electric dipole moments, so Stark shifts are second-order effects. From second-order perturbation theory, they are therefore inversely proportional to the energy difference of states of opposite parity. Since rotational energies are much smaller than typical atomic energies, these shifts can be larger in molecules. For molecules with internal electronic angular momentum, even small electric fields can quickly become nonperturbative and can polarize the molecule~\cite{carrNJP2009}. This effect arises because there are two projections of this angular momentum on the internuclear axis. The projections combine to form parity eigenstates that are nearly degenerate. These so-called $\Lambda$- or $\Omega$-doublets are split only due to a Coriolis coupling with the overall rotation of the molecule.

In order to suppress these potentially large Stark shifts, several experiments use homonuclear molecules. In such molecules, nuclear exchange symmetry forces all rotational states to have the same parity, which eliminates half the rotational states and makes the molecule nonpolar. Since the closest states of opposite parity are excited electronic states, the Stark shifts are suppressed. Thus, homonuclear molecules should have relatively small sensitivity to stray electric fields, to the oscillating electric fields of ion traps, and to blackbody radiation. Nonetheless, AC Stark shifts from applied laser radiation are still expected to be a major source of systematic uncertainties~\cite{kondovNatPhys2019,carolloAtoms2018,wolfArxiv2020}. For experiments in an optical lattice, it will be important to tune the lattice laser wavelength such that there is no differential AC Stark shift between the two clock states~\cite{zelevinskyPRL2008,kondovNatPhys2019}. For polar molecules, the Stark shifts are in general not small. There are some cases, however, where the scalar polarizability cancels, leaving only the tensor polarizability. In these cases, it should be possible to null Stark shifts by averaging over multiple transitions with different $M$ and $M^\prime$ quantum numbers~\cite{kokishPRA2018}.

Magnetic field (Zeeman) shifts appear with any molecular magnetic dipole moment. States with electronic angular momentum often have magnetic dipole moments of order a Bohr magneton $\mu_B$ and shifts of order $\mu_B/h = 1.4\times10^{10}~{\rm Hz/T}$. These shifts can be suppressed by around $10^3$ if the spin and orbital magnetic moments cancel ($\Lambda = -2\Sigma$), such as in the states $^2\Pi_{1/2}$ and $^3\Delta_1$. Magnetic dipoles from the molecule's rotation and nuclear spins are typically smaller than $\mu_B$ by of order $10^3$ because of the heavier nuclear mass~\cite{brown_carrington}. Linear Zeeman shifts can be nulled in a transition by driving $M=0\leftrightarrow M^\prime=0$ or (when such states do not exist) by calculating the average of $M\leftrightarrow M^\prime$ and $-M\leftrightarrow -M^\prime$. Second-order Zeeman shifts can arise from mixing with other states. After calibrating the field-dependence of this shift, it can be calculated by measuring the value of the magnetic field with a first-order-shifted transition.

Doppler shifts arise from relative motion of the molecule and laser source. First-order Doppler shifts are suppressed by tight confinement in a trap, which also facilitates long probe times. Residual first-order shifts, such as relative motion within an apparatus, can be monitored by probing the molecules from different directions. Second order Doppler shifts -- a relativistic time dilation effect -- arise from nonzero mean-squared speed, such as from finite temperature or micromotion in an ion trap. This is the largest systematic uncertainty in ion-based atomic clocks (at $6\times10^{-19}$,~\cite{brewerPRL2019}) and is expected to be a limiting uncertainty in ion-based molecular clocks as well.

Among other shifts, the quadrupolar electric field in ion traps will interact with molecular states that have an electric quadrupole moment. States with $J=0$ or $\tfrac{1}{2}$ have no such moment, so the shift is exactly zero. Collisional/density shifts could play a role at high densities in 1D optical lattices. General relativity causes a gravitational redshift of $10^{-18}$ at 1~cm height differences. This will be important if the reference clock is not at the same location (and height) as the molecule. Implementation-specific shifts will arise as well. While the largest systematic effect will depend on the particular molecule and experiment, several estimates have shown the potential for uncertainties on the shifts to be of order $10^{-18}$~\cite{kajitaPRA2014,schillerPRL2014,kajitaPRA2017,carolloAtoms2018,kokishPRA2018}, comparable to those of optical atomic clocks.

\subsection{Statistical uncertainties} \label{sec:stats}
Statistical uncertainties are limited by the number and duration of experiments. An order-of-magnitude estimate of the statistical uncertainty $\delta f$ of a frequency measurement is given by~\cite{ludlowRMP2015}
\begin{equation}
	\delta f \sim \frac{1}{\sqrt{N T_m \tau}}, \label{eq:deltaf}
\end{equation}
where $N$ is the number of molecules probed, $T_m$ is the duration of a single measurement, and $\tau$ is the total measurement duration. The precise uncertainty will depend on technical details, such as the pulse sequence employed. \Eref{eq:deltaf} assumes negligible time is dedicated to state preparation, measurement, and pulse durations in any Ramsey sequence. It further assumes that the statistical uncertainty is limited by quantum projection noise~\cite{itanoPRA1993}. For transitions with extremely narrow natural linewidths, such as vibrational overtones of nonpolar molecules, the probe time $T_m$ will be limited by technical considerations like the laser linewidth. For polar molecules, it may be limited by the transition's natural linewidth.

Given this frequency uncertainty, the fractional uncertainty in $\mu$ is 
\begin{equation}
	\frac{\delta\mu}{\mu} = \frac{\delta f}{f_\mu} = \frac{\delta f}{K_\mu f}.
\end{equation}
 As an example, consider a vibrational transition with $K_\mu = -0.5$ probed by a laser of wavelength 1~$\mu$m ($f=300$~THz) for a duration of $T_m = 1$~s. It would achieve a statistical uncertainty $\delta\mu/\mu \sim 7\times10^{-15}/\sqrt{N(\tau/{\rm s})}$, which would average down to the present limit of $6.5\times10^{-17}$ in around three hours with one molecule.

For probes of oscillating constants by use of dynamical decoupling, the statistical uncertainty in \eref{eq:deltaf} is still the correct order of magnitude~\cite{kotlerPRL2013}. In a such a multi-pulse sequence, other technical considerations may come into play. For example, the pulse fidelity may limit the number of pulses and thus the duration of an individual experiment $T_m$. As the search frequency increases, the finite pulse duration will need to be included in the filter-function design. The overall scaling will change if either $T_m$ or $\tau$ approach the dark matter coherence time $\tau_{\rm coh}$~\cite{dereviankoPRA2018,kotlerPRL2013,degenRMP2017}.

Searches for oscillations by use of traditional spectroscopy techniques~\cite{demtroderBook} will have statistical limits based on quantities such as shot noise in a photodetector. For a given signal $S$ and noise $\delta S$, one can achieve a fractional frequency uncertainty of~\cite{ludlowRMP2015}
\begin{equation}
	\delta f = \frac{\delta S}{dS/df}.
\end{equation}
Here, $dS/df$ is a discriminator that says how much your signal changes for a given frequency shift. This slope is typically steeper for narrower resonances, such that we can approximate the statistical uncertainty by
\begin{equation}
	\delta f \sim \frac{\delta S}{S/\gamma} = \frac{\delta S}{S}\frac{f}{Q},
\end{equation}
where $S/\delta S$ is the signal-to-noise ratio and $Q=f/\gamma$ is the quality factor. For shot-noise-limited experiments, $\delta S/S = \sqrt{e/(I\tau)}$, where $e$ is the magnitude of the electron charge, $I$ is the average detector current, and $\tau$ is the integration time. For example, a shot-noise-limited experiment with $Q=10^8$ (such as $\gamma=3$~MHz, $f=300$~THz) and $I=1$~mA of average current would have a statistical limit of $\delta f/f \sim 1\times10^{-16}/\sqrt{\tau/{\rm s}}$. \Eref{eq:linewidthSensitivity} converts this frequency limit to $\delta \mu/\mu$ by use of the relative sensitivity $K_\mu$ and the decaying response for $f_\phi\gtrsim\gamma$. For $f_\phi\ll\gamma$ and $K_\mu=-0.5$, the same example's shot-noise limit is $\delta\mu/\mu\sim 2\times10^{-16}/\sqrt{\tau/{\rm s}}$. In a realistic experiment, laser technical noise may dominate over shot noise at low frequencies~\cite{antypasPRL2019}. Because the molecules cannot effectively track oscillations with $f_\phi\gg\gamma$, the statistical sensitivity to $\mu$ at high frequencies becomes modulation-frequency-dependent. Keeping the same example, it is $\delta\mu/\mu\sim8\times10^{-23}(f_\phi/{\rm Hz})/\sqrt{\tau/{\rm s}}$. Note that this fast-modulation result is actually independent of $\gamma$.

\section{Conclusion}

Molecular vibrations are leading systems for next-generation searches for drifts in $\mu$ and for the first-ever direct searches for oscillations in $\mu$. The absolute sensitivity of vibrational frequencies to $\mu$ can be estimated with a relatively simple anharmonic oscillator model. Several molecules possess optical-frequency vibrational overtones with potential instabilities at the $10^{-18}$ level or below, comparable to the best atomic clocks. Experiments are underway with a variety of molecules, with most currently refining the state-control techniques. Future results will discover or further constrain the couplings and masses of new fields, with implications for quantum gravity and dark matter.

\ack
We thank Roee Ozeri and Joshua Eby for discussions. This research was funded by the U.S. National Science Foundation Grant PHY-1806223.

\end{document}